# Field Effect Transistors for Terahertz Detection: Physics and First Imaging Applications

**W. Knap, M. Dyakonov, D. Coquillat, F. Teppe, N. Dyakonova**

*Université Montpellier2 - CNRS, Place E. Bataillon, 34950 Montpellier, France*

**J. Łusakowski, K. Karpierz, M. Sakowicz**

*Institute of Experimental Physics, University of Warsaw, ul. Hoża 69, 00-681 Warsaw, Poland*

**G. Valusis, D. Seliuta, I. Kasalynas**

*Semiconductor Physics Institute, A. Gostauto 11, LT-01108 Vilnius, Lithuania*

**A. El Fatimy, Y.Meziani, T. Otsuji**

*Tohoku University RIEC Ultra-broadband Signal Processing 2-1-1 Katahira, Aoba-ku 980-8577 Japan*

**Abstract:** Resonant frequencies of the two-dimensional plasma in FETs increase with the reduction of the channel dimensions and can reach the THz range for sub-micron gate lengths. Nonlinear properties of the electron plasma in the transistor channel can be used for the detection and mixing of THz frequencies. At cryogenic temperatures resonant and gate voltage tunable detection related to plasma waves resonances, is observed. At room temperature, when plasma oscillations are overdamped, the FET can operate as an efficient broadband THz detector. We present the main theoretical and experimental results on THz detection by FETs in the context of their possible application for THz imaging.

1. Introduction

The channel of a field effect transistor (FET) can act as a resonator for plasma waves. The plasma frequency of this resonator depends on its dimensions and for gate lengths of a micron and sub-micron (nanometer) size can reach the terahertz (THz) range. The interest in the applications of FETs for THz spectroscopy started at the beginning of '90s with the pioneering theoretical work of Dyakonov and Shur [1] who predicted that a steady current flow in a FET channel can become unstable against generation of the plasma waves. These waves can, in turn, lead to the emission of the electromagnetic radiation at the plasma wave frequency. This work was followed by the another one where the same authors have shown that the nonlinear properties of the 2D plasma in the transistor channel can be used for detection and mixing of THz radiation [2]. It is worth noting that this work treated rigorously and gave a complete description of the resonant as well as the non-resonant (overdamped) plasma oscillation regimes.

THz emission in the nW power range from submicron GaAs and GaN FETs has been observed both at cryogenic as well as at room temperatures [3-5]. At the moment, however, FET based THz microsources can not compete with existing Quantum Cascade Lasers (QCL) or Time Domain Spectroscopy (TDS) sources in the practical applications. It appeared, nevertheless, that THz detection by FETs can be very promising and close to applications.

Here we present an overview of the recent results on detection obtained in different types of III-V semiconductor-based nanometer-sized High Electron Mobility Transistors (HEMTs) [6-12]. Many experimental results were obtained at cryogenic temperatures where the resonant plasma modes can be excited [8, 9, 11, 12]. However, already in the first experiments, it was shown that GaAs/AlGaAs and GaInAs/GaAs HEMTs can also operate as THz broadband detectors at room temperatures [6, 7, 9]. From the application point of view research on Si-MOSFETs was very important. It has been demonstrated that Si-MOSFETs can be efficient room temperature detectors of sub-THz radiation and can be used up to 2.5 THz as well [13, 14]. Their noise equivalent power was found to be one of the lowest of all room temperature operating fast THz detectors [14]. They have been recently integrated in focal plane arrays and checked for imaging applications [15, 16].

The main well established facts about THz detection by FETs are: i) the resonant detection observed at cryogenic temperatures is due to plasma waves related rectification and ii) at room temperature the plasma wave oscillations are overdamped but the rectification mechanism is still efficient and enables a broadband THz detection and imaging.

The paper is organized as follows. Section 2 describes the basic principles of the detection with FETs. Section 3 presents an overview of the main experimental results on sub-THz and THz detection. Recent results on plasma resonance narrowing by geometrical factors and the drain current are also presented therein. Section 4 gives practical examples concerning THz imaging.

## 2. Principles of terahertz detection by FETs

The idea of using a FET for emission and detection of THz radiation was put forward by Dyakonov and Shur [1, 2]. The possibility of the detection is due to nonlinear properties of the transistor, which lead to the rectification of an ac current induced by the incoming radiation. As a result, a photoresponse appears in the form of dc voltage between source and drain which is proportional to the radiation power (photovoltaic effect). Obviously, some asymmetry between the source and drain is needed to induce such a voltage.

There may be various reasons of such an asymmetry. One of them is the difference in the source and drain boundary conditions due to some external (parasitic) capacitances. Another one is the asymmetry in feeding the incoming radiation, which can be achieved either by using a special antenna, or by an asymmetric design of the source and drain contact pads. Thus the radiation may predominantly create an ac voltage between the source and the gate (or between the drain and the gate) pair of contacts. Finally, the asymmetry can naturally arise if a dc current is passed between source and drain, creating a depletion of the electron density on the drain side of the channel [3]. In most of the experiments carried out so far, the THz radiation was applied to the transistor channel, together with contact pads and bonding wires. In such a case, it is obviously difficult to define how exactly the radiation is coupled to the transistor. Important experimental information about the way of coupling to a real device was obtained from experiments with polarized sub-THz radiation [17, 18] and with focused THz radiation [19]. These results are presented in Sect. 3. Theoretically, we will consider the case of an extreme asymmetry, where the incoming radiation creates an ac voltage with amplitude $U_a$ only between the source and the gate, see Fig. 1. We will also assume that there is no dc current between the source and drain.

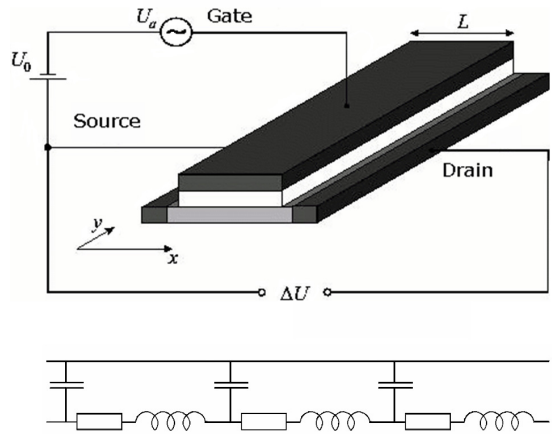

Fig. 1. Schematics of a FET as a THz detector (above) and the equivalent circuit (below).

Generally, the FET may be described by an equivalent circuit presented in Fig. 1. The obvious elements are the distributed gate-to-channel capacitance and the channel resistance, which depends on the gate voltage through the electron concentration in the channel:

$$en = CU, \qquad (1)$$

where $e$ is the elementary charge, $n$ is the electron concentration in the channel, $C$ is the gate-to-channel capacitance per unit area, and $U$ is the gate to channel voltage. Note, that Eq. 1 is valid *locally*, so as long as the scale of the spatial variation of $U(x)$ is larger than the gate-to-channel separation (the gradual channel approximation). Under static conditions and in the absence of the drain current, $U = U_0 = V_g - V_{th}$, where $U_0$ is the voltage swing, $V_g$ is the gate voltage, and $V_{th}$ is the threshold voltage at which the channel is completely depleted. The inductances in Fig. 1 represent so called *kinetic* inductances, which are due to the electron inertia and are proportional to $m$, the electron effective mass. Depending on the frequency $\omega$, one can distinguish two regimes of operation, and each of them can be further divided into two sub-regimes depending on the gate length $L$.

1. *High frequency regime* occurs when $\omega\tau > 1$, where $\tau$ is the electron momentum relaxation time, determining the conductivity in the channel $\sigma = ne^2\tau/m$. In this case, the kinetic inductances in Fig. 1 are of primordial importance, and the plasma waves analogous to the waves in an RLC transmission line, will be excited. The plasma waves have a velocity $s=(eU/m)^{1/2}$ [1] and a damping time $\tau$. Thus their propagation distance is $s\tau$.

1*a. Short gate, $L < s\tau$*. The plasma wave reaches the drain side of the channel, gets reflected, and forms a standing wave with an enhanced amplitude, so that the channel serves as a resonator for plasma oscillations. The fundamental mode has the frequency $\sim s/L$, with a numerical coefficient depending on the boundary conditions.

1*b. Long gate, $L>>s\tau$*. The plasma waves excited at the source will decay before reaching the drain, so that the ac current will exist only in a small part of the channel adjacent to the source.

2. *Low frequency regime, $\omega\tau <<1$*. Now, the plasma waves cannot exist because of an overdamping. At these low frequencies, the inductances in Fig. 1 become simply short-circuits which leads to an RC line. Its properties further depend on the gate length, the relevant parameter being $\omega\tau_{RC}$, where $\tau_{RC}$ is the RC time constant of the whole transistor. Since the total channel resistance is $L\rho/W$, and the total capacitance is $CWL$ (where $W$ is the gate width and $\rho = 1/\sigma$ is the channel resistivity), one finds $\tau_{RC} = L^2\rho C$.

2*a. Short gate, $L < (\rho C\omega)^{-1/2}$*. This means that $\omega\tau_{RC} <1$, so that the ac current goes through the gate-to-channel capacitance practically uniformly on the whole length of the gate. This is the so-called "resistive mixer" regime [20-22]. For the THz frequencies this regime can apply only for transistors with extremely short gates (smaller than 70 nm at 1 THz in silicon).

2*b. Long gate*, $L >> (\rho C\omega)^{-1/2}$. Now $\omega\tau_{RC} >>1$, and the induced ac current will leak to the gate at a small distance $l$ from the source, such that the resistance $R(l)$ and the capacitance $C(l)$ of this piece of the transistor channel satisfy the condition $\omega\tau_{RC}(l) = 1$, where $\tau_{RC}(l) = R(l)C(l) = l^2\rho C$. This condition gives the value of the "leakage length" $l$ on the order of $(\rho C\omega)^{-1/2}$ (which can also be rewritten as $s(\tau/\omega)^{1/2}$). If $l<< L$, then neither ac voltage, nor ac current will exist in the channel at distances beyond $l$ from the source, see Fig. 2.

Thus, the characteristic length where the ac current exists is $s\tau$ for $\omega\tau >1$, and $s(\tau/\omega)^{1/2}$ for $\omega\tau <1$ [2]. Let us now present few quantitative examples for the different cases presented above. For $\tau = 30$ fs ($\mu = 300$ cm$^2$/(Vs) in Si MOSFET) and $s = 10^8$ cm/s the regime 1 will be realized for the radiation frequencies, $f$ greater then 5 THz; the regime 1*a* for $L < 30$ nm. For $f = 0.5$ THz (the regime 2), one finds the characteristic gate length distinguishing regimes 2*a* and 2*b* to be around 0.1 μm. If the conditions of the case 1*a* are satisfied, the photoresponse will be resonant, corresponding to the excitation of discrete plasma oscillation modes in the channel. Otherwise, the FET will operate as a broad-band detector.

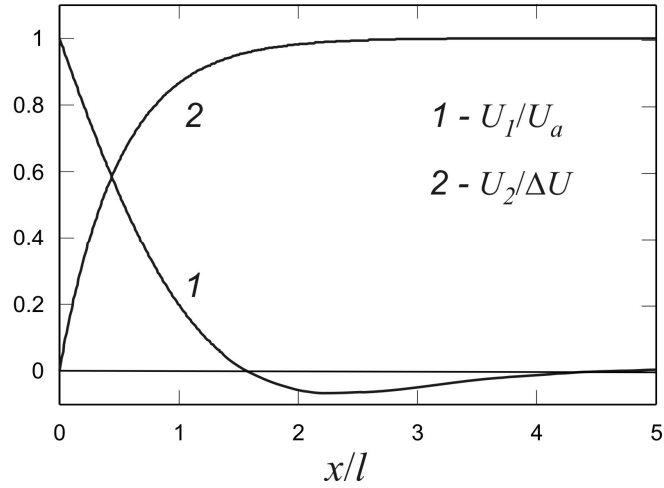

Fig. 2. Dependence of the ac voltage $U_1/U_a$ at $\omega t = 2\pi n$ and of the dc photoinduced voltage $U_2/\Delta U$ on the distance from the source $x$ for a long gate.

For a long gate, there is no qualitative difference between the low-frequency regime ($\omega\tau \ll 1$), when plasma waves do not exist (the case 2b) and the high frequency regime ($\omega\tau \gg 1$), where plasma oscillations are excited (the case 1b). There is however some quantitative differences, see Eq. 2 below. Anyway, plasma waves are excited and their existence in this case has been clearly confirmed by the recent detection experiments in the magnetic fields [23]. The plasma waves cannot propagate above the cyclotron frequency. Therefore, in experiments with a fixed radiation frequency the photoresponse is strongly reduced when the magnetic field goes through the cyclotron resonance. This is probably the most spectacular manifestation of the importance of plasma waves in the terahertz detection by FETs [23].

*Mechanism of the nonlinearity.* The most important mechanism is the modulation of the electron concentration in the channel, and hence of the channel resistance, by the local ac gate-to-channel voltage, as described by Eq. 1. Because of this, in the expression for the electric current $j = env$, both the concentration, $n$ and the drift velocity, $v$ will be modulated at the radiation frequency. As a result, a dc current will appear: $j_{DC} = e\langle n_1(t)v_1(t)\rangle$, where $n_1(t)$ and $v_1(t)$ are the modulated components of $n$ and $v$, and the angular brackets denote averaging over the oscillation period $2\pi/\omega$. Under open circuit conditions a compensating dc electric field will arise, resulting in the photoinduced source-drain voltage $\Delta U$.

*A simplified theory.* The most important case is that of a long gate (the regimes 1b and 2b) when, independently of the value of the parameter $\omega\tau$, the ac current excited by the incoming radiation at the source cannot reach the drain side of the channel. For this case within the hydrodynamic approach the following result for the photoinduced voltage was derived [2]:

$$\Delta U = \frac{U_a^2}{4U_0}\left(1 + \frac{2\omega\tau}{\sqrt{1+(\omega\tau)^2}}\right). \quad (2)$$

As seen from this formula, the photoresponse changes only by a factor of 3, as the parameter $\omega\tau$ increases from low to high values, even though the physics becomes different: at $\omega\tau>1$ plasma waves are excited, while at $\omega\tau<1$ they are not.

The basic equations may be written as [1, 2]:

$$\frac{\partial U}{\partial t} + \frac{\partial}{\partial x}(Uv) = 0, \tag{3}$$

$$\frac{\partial v}{\partial t} = -\frac{e}{m}\frac{\partial U}{\partial x} - \frac{v}{\tau}. \tag{4}$$

Here Eq. 3 is the continuity equation, in which the concentration $n$ is replaced by $U$ using Eq. 1, while Eq. 4 is the Drude equation for the drift velocity $v$ [24]. The boundary condition for gate-to-channel voltage at the source side of the channel $(x = 0)$ is: $U(0,t) = U_0 + U_a \cos(\omega t)$. For a long gate, the boundary condition at the drain is $v(\infty) = 0$.

The inertial term $\partial v/\partial t$ is accounted for by the kinetic inductances in Fig. 1. Here, we will consider only the simple case $\omega\tau<1$, when the inertial term can be neglected. Then $v = -\mu \partial U/\partial x$, and

$$\frac{\partial U}{\partial t} = \mu \frac{\partial}{\partial x}\left(U \frac{\partial U}{\partial x}\right), \tag{5}$$

where $\mu = e\tau/m$ is the electron mobility.

We search the solution of the nonlinear Eq. 5 as an expansion in powers of $U_a$: $U = U_0 + U_1 + U_2$, $U_1$ is the ac voltage, proportional to $U_a$, and $U_2$ is the time-independent contribution proportional to $U_a^2$ (the photovoltage). In the first order in $U_a$ we obtain the diffusion equation for $U_1$ [25]:

$$\frac{\partial U_1}{\partial t} = s^2 \tau \frac{\partial^2 U_1}{\partial x^2}, \tag{6}$$

with the boundary conditions $U_1(0,t) = U_a \cos(\omega t)$, $U_1(\infty,t) = 0$. The solution of this equation is:

$$U_1(x,t) = U_a \exp(-x/l) \cos(\omega t - x/l), \tag{7}$$

where the characteristic length $l$ for the decay of the ac voltage (and current) away from the source is given by:

$$l = s(2\tau/\omega)^{1/2}. \tag{8}$$

This length defines the size of the part of the transistor adjacent to the source, whose resistance and the capacitance are such that $\omega\tau_{RC}(l) \sim 1$, as explained above.

In the second order in $U_a$, Eq. 5 yields:

$$U_0 \frac{\partial U_2}{\partial x} + \left\langle U_1 \frac{\partial U_1}{\partial x} \right\rangle = 0, \tag{9}$$

which means simply the absence of the dc current. Integrating this equation, one obtains:

$$U_2(x) = \frac{1}{2U_0}\left[\langle U_1^2(0,t)\rangle - \langle U_1^2(x,t)\rangle\right], \qquad (10)$$

where the time averaged quantity $\langle U_1^2(x,t)\rangle = (1/2)U_a^2\exp(-2x/l)$ is found from Eq. 7. Thus, the photovoltage $\Delta U = U_2(\infty)$ coincides with Eq. 2, provided that $\omega\tau \ll 1$. Figure 2 shows the ac voltage $U_1$ and the build-up of the dc voltage $U_2$ as functions of the distance from the source.

The maximum photovoltage is achieved at $U_0$ close to 0 V, where the relative ac modulation of the electron concentration in the channel is the strongest (note that Eq. 1 is not valid in the vicinity of $U_0 = 0$ V).

It is instructive to compare the FET detector with the well known Schottky diode detector. In both devices the detection process is based on a rectification of the incident THz field by a nonlinear element. However, there are some important differences. The nonlinearity in the Schottky diode is due to the nonlinear I-V characteristic of the potential barrier between the metal and the semiconductor. The physical origin of the nonlinearity in the case of the FET transistor is very different. As discussed above, it is due to the fact that the incident THz radiation modulates both the carrier drift velocity and the carrier density. The static I-V dependence has no direct relevance to the detection properties of the FET.

### 3. Experiments on THz detection with FETs

The layout of a typical commercial device used in experiments is shown in Fig. 3. Usually, the device is mounted on a standard DIL package. In most experiments, transistors are illuminated by a monochromatic beam generated by either a Gunn diode, a backward wave oscillator (BWO), or a $CO_2$ pumped molecular THz laser. In the absence of an antenna, the amount of THz radiation that is coupled to the device is probably very small. Therefore a big progress in the sensitivity can be obtained by adding a proper antenna and/or a cavity coupling. The photoresponse as a function of the gate voltage was measured and the plasma resonances have been identified by their gate voltage and temperature dependencies. Possible artifacts due to resonances related to radiation coupling were carefully investigated and ruled out by experiments with different angles of illumination.

As can be seen in Fig. 3, the geometry of metallization is such that the capacitance between the source and the gate is larger than the capacitance between the drain and the gate contacts. This asymmetry defines the sign of the signal for a uniform device illumination. Experiments with polarized radiation usually show a well defined preferential orientation of the electric field of incoming THz radiation. This orientation is defined by geometry of metallization of the contacts or by the bonding wires [17, 18]. The black solid line in Fig. 3 shows the amplitude of the photoresponce versus polarization angle obtained with a linearly polarized radiation. The maximum in the signal is observed when the polarization is such that the electric field is along the direction from the drain contact to the gate contact. This suggests that the ac voltage is predominantly created between the drain and the gate contacts.

From the point of view of future applications, it is important to optimize the radiation-detector coupling. This problem was recently addressed by Sakowicz *et al.*

[17, 18]. It was found that for relatively low sub-THz frequencies ~100 GHz, the radiation couples to the transistor via an antenna formed mainly by the bonding wires [17]. For frequencies above 200 GHz, these are the metal pads of the source, drain, and gate contacts that play the role of an antenna, and the radiation coupling efficiency is strongly influenced by the layout of these electrodes [18] (see Fig. 3 and discussion above).

The mechanism of the radiation coupling was also studied in Ref. 19. In the case of a fully symmetric transistor, the photovoltaic signal appeared only if the THz illumination was asymmetric. The maximum of the signal was obtained when the beam was focused outside of the transistor structure close to the drain or close to the source contact pads. The photovoltaic signal was not observed when the beam was focalized on the central part of the transistor.

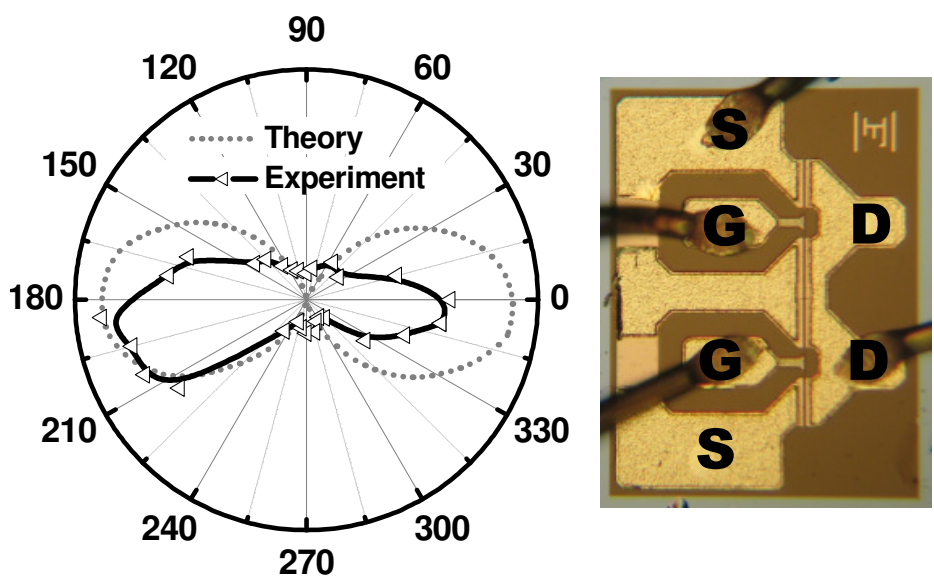

Fig. 3. The amplitude of the photovoltaic signal for 285 GHz versus polarization angle and the layout of a typical commercial FET device used in experiments. The solid line corresponds to experimental results, and the dotted line is the theoretical dependence for the preferential polarization along the gate - drain direction [18]. S, G and D denote source, gate and drain, respectively.

Recently new results on the radiation coupling mechanism have been presented by Tanigawa *et al.* [26]. They integrated a patch antenna onto GaN HEMT and showed that it can enhance the detection sensitivity by two orders of magnitude. It was also found that the optimal polarization is defined by the orientation of the antenna, and not by the channel orientation.

Typical results of detection experiments are shown in Fig. 4 [9]. A photovoltaic signal between source and drain is recorded versus the gate voltage, i.e. versus the carrier density in the channel. In Fig. 4a, the registered signal is presented for two transistors with different values of threshold voltage. For a high carrier density (open channel) the signal is relatively small and increases when the gate voltage approaches the threshold. This increase follows the $1/(V_g - V_{th})$ functional dependence in agreement with Eq. 2. However, the signal does not diverge when $V_g$ approaches $V_{th}$ but it usually shows a broad maximum. Comparing the photovoltaic signal (Fig. 4a) and the transfer characteristics (Fig. 4b) one can notice that the position of this maximum is correlated with the threshold voltage. The shape and the position of the maximum depend also on loading effects [27]. Close to the threshold voltage, the channel resistance diverges to infinity and the transistor behaves like a voltage source with a very high internal resistance. The signal limitation close to the transistor threshold voltage can be also due to gate leakage currents [9]. The detection curves like shown in Fig. 4a are related to a broadband nonresonant detection described as the cases 1*b*, 2*a* or 2*b* in the Sect. 2 (theoretical part). The position and the shape of the maxima are also strongly temperature dependent because the carrier density (threshold voltage), the channel resistance and the gate leakage currents depend strongly on the temperature. This is illustrated in Fig. 4c where results for the same transistor at three different temperatures are shown. The maximum of the detection shifts to the lower gate voltage with decrease in temperature. For the lowest temperature one can observe an additional maximum appearing on the $1/(V_g - V_{th})$ like shoulder. This maximum is a signature of the resonant detection.

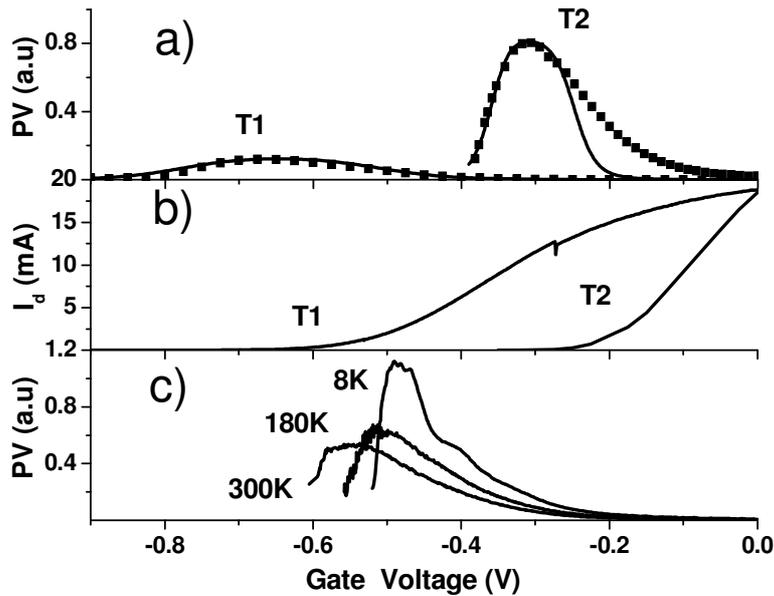

Fig. 4. a) The experimental photoresponse for GaAs/AlGaAs FETs. b) The drain current versus the gate voltage $V_g$ for two transistors T1 and T2. Curves marked T1 correspond to the transistor with the threshold voltage $V_{th}$ = - 0.55 V measured at 300 K and at the frequency 200 GHz. Curves marked T2 correspond to another transistor with threshold voltage $V_{th}$ = - 0.22 V measured at 210 K and at 100 GHz. b) Drain current versus gate voltage for the transistors T1 and T2. c) The photoresponse of the transistor with the gate length $L$ = 0.15 µm as a function of the gate voltage at the radiation frequency of 600 GHz, the threshold changes with temperature. At 8 K a resonant structure appears.

Examples of a resonant detection (the case 1*a* described above), are shown in Fig. 5 and Fig. 6. We show experimental traces as well as theoretical fits for the gate voltages bigger then the threshold where the measurement system related artifacts (resistive, capacitive coupling and gate leakage problems) can be neglected. In the majority of experiments, the incoming radiation is a monochromatic beam and the source-drain voltage is recorded versus the gate voltage. The gate voltage controls the carrier density in the channel and therefore allows the resonant plasma frequency to be tuned. A resonant enhancement of the registered voltage is observed once the resonant plasma frequency coincides with the frequency of the incoming THz radiation. The resonance appears in the lowest temperatures because the carrier mobility increases and $\omega\tau > 1$ condition can be reached. Figure 5b shows results of fits according the theory of Ref. 2. The mobility shown in the inset to Fig. 5b was measured in experiments described in Ref. 8. One can see that both the slope of the detected signal as well as appearance of the resonance can be well reproduced.

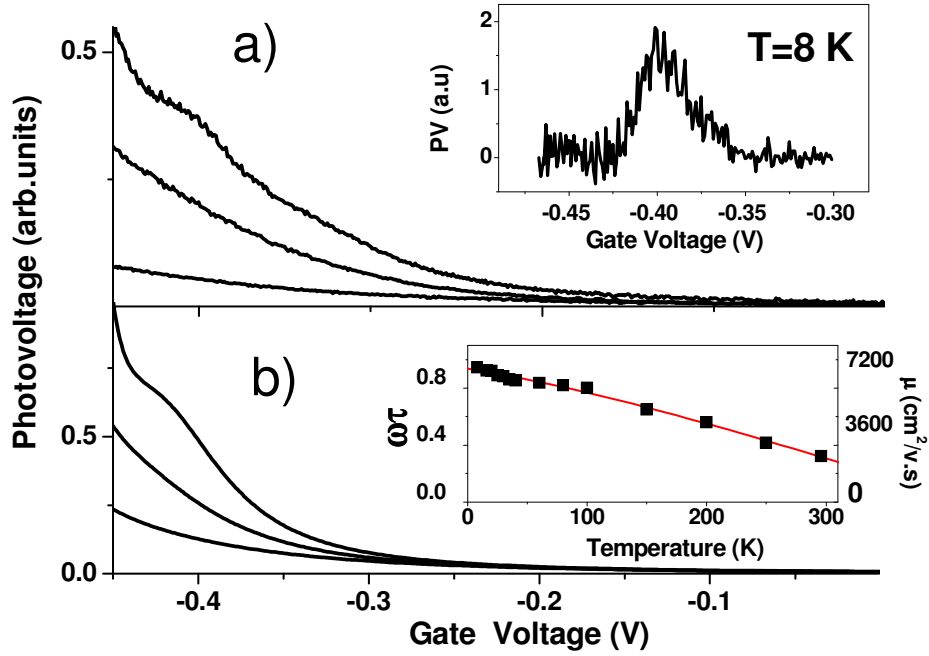

Fig. 5. Experimental (a) and theoretical (b) curves of the photoresponse versus the gate voltage in the region above the threshold. An evolution of the photoresponse with the decrease of temperature (300 K, 180 K, 8 K from the bottom to the top curve) and the appearance of the resonant structure are illustrated. Inset in a) shows the resonant signal after subtraction of the background. Inset in b) shows the evolution of the carrier mobility and the corresponding quality factor for 600 GHz radiation [8, 10].

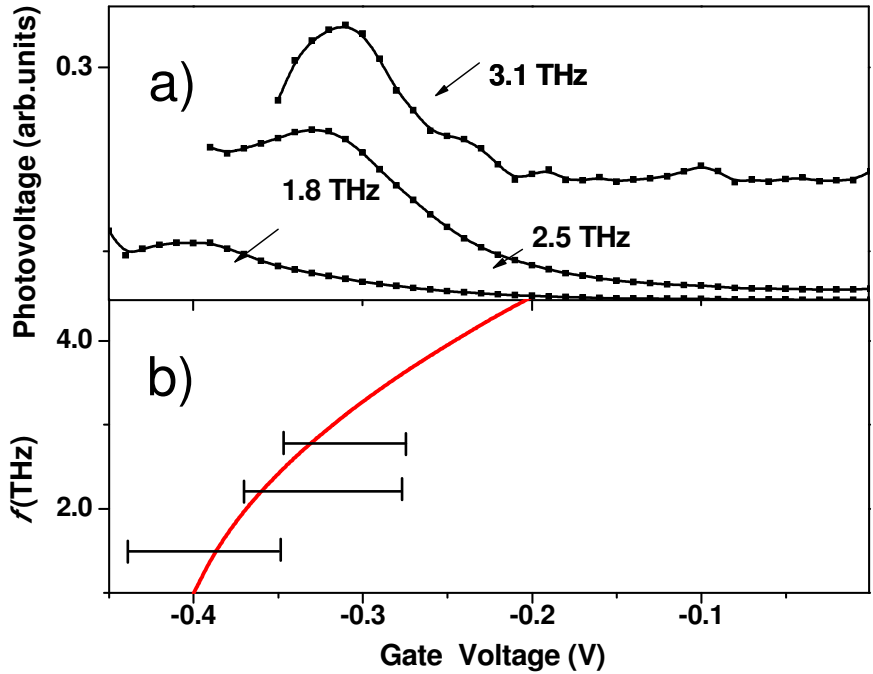

Fig. 6. a) Resonant response of a high mobility InGaAs/InAlAs transistors for 1.8 THz, 2.5 THz and 3.1 THz registered at 10 K. b) Position of the maxima versus the gate voltage (bars) together with the theoretical calculation (solid line) - after Ref. 11.

First resonant detection experiments were carried out on submicron GaAs/AlGaAs HEMTs [6-9]. Subsequently, high mobility InGaAs/InAlAs transistors were studied [11]. Figure 6a shows an example of a plasma related resonant detection at 1.8 THz, 2.5 THz and 3.1 THz for InGaAs/InAlAs transistors registered at 10 K. One can see the resonant detection that corresponds to the case 1a, when $\omega\tau > 1$ and the gate is short enough, see the discussion in Sect. 2. In Fig. 6b the position of the resonant maxima is shown. As the excitation frequency increases from 1.8 THz to 3.1 THz, the plasma resonance moves to higher swing voltage in an approximate agreement with the theoretical predictions (solid line). However, the resonance is much broader than theoretically expected.

A significant plasma resonance broadening appears to be one of the main unresolved problems of the resonant THz detection by FETs. The most convenient way to discuss the broadening is to consider the quality factor $Q$. From the experimental point of view, $Q$ is an important quantity because it describes the ratio of the resonant line position (in frequency or voltage) with respect to the line width. Theoretically, the quality factor can be written as $Q = \omega\tau$, where $\omega = 2\pi f$ is the excitation frequency and $\tau$ is the electron momentum relaxation time. The inverse of $\tau$ corresponds to the plasma resonance linewidth in units of frequency.

For $\omega\tau \ll 1$, the plasma oscillations are overdamped and consequently, the response is expected to show a non resonant, monotonic behavior. In the opposite case, when the quality factor becomes much larger than 1, narrow plasma resonance peaks are expected. The main motivation behind changing the transistor material system from GaAs/GaAlAs, as used in the first experiments [3, 6], to InGaAs/InP [11] and using higher excitation frequencies (up to 3 THz instead of 0.6 THz) was to improve the quality factor.

A longer carrier scattering time (a higher mobility) combined with the use of higher excitation frequencies was expected to result in an increase of the quality factor by at least an order of magnitude, leading to sharp plasma resonances. However, experimentally observed plasma resonances remain broad. Even at 3 THz excitation the quality factor $Q$ is never higher than 2–3. Understanding the origin of the broadening and minimizing it is one of the most important experimental and theoretical challenges.

Two main hypotheses of the origin of an additional broadening are currently under consideration: i) existence of oblique plasma modes [28] and ii) an additional damping due to the leakage of gated plasmons to ungated parts of the transistor channel [29]. The first hypothesis is related to the fact that in realistic devices the gate width is much greater than the gate length. Thus, the transistor channel serves as a waveguide rather than a resonator for plasma waves. In such a case, plasma waves can propagate not only in the source-drain direction but also in oblique directions. The spectrum of plasma waves in this case is continuous, with a cut-off at low frequency. The second hypothesis, a leakage of gated plasmons to ungated parts of the channel [29], is related to the observation that normally the gate covers only a small part of the source-drain distance. Therefore, the plasma under the gate can not be treated independently of the plasma in ungated parts. An interaction between the two plasma regions can lead not only to a modification of the resonant frequency [30, 31], but also to line broadening.

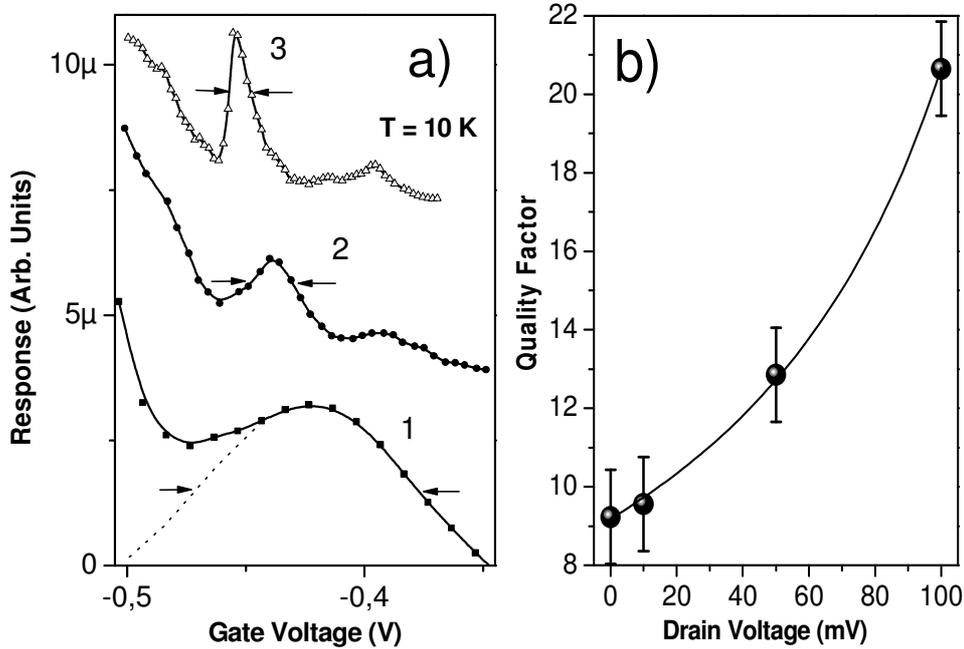

Fig. 7. a) Photoresponse at 10 K as a function of the gate voltage. Curve 1 was obtained at 2.5 THz with a conventional InGaAs HEMT. Curves 2 and 3 were obtained at 540 GHz with a multi-channel InGaAs HEMT at two different drain-to-source voltages (0 mV and 100 mV, respectively). Curves are vertically shifted. b) Quality factor of the resonance as a function of the source-drain voltage. The solid line is a fit of experimental data using Eq. 13.

To decrease the role of the oblique modes, one has to change the geometry of the channel. One can, for example replace the wide single channel by a series of many narrow channels [32, 33]. In Fig. 7a we present a comparison of the photoresponse at 2.5 THz of a conventional InGaAs HEMT (curve 1, squares), with that of a narrow multichannel transistor at 540 GHz (curve 2, circles). One can see the narrowing of the resonant line, even though a lower incident frequency was used.

Another way to decrease the broadening of the plasma resonances is to apply a drain current. The drain current affects the plasma relaxation rate by driving the two-dimensional plasma in the transistor channel towards the Dyakonov-Shur plasma wave instability. For $\omega\tau \gg 1$, the FET operates as a resonant detector. In this case, the induced photoresponse is given by [2]:

$$\Delta U \propto \frac{1}{(\omega-\omega_0)^2 + \left(\frac{1}{2\tau}\right)^2} \quad (11)$$

where $\omega_0$ is the fundamental resonant plasma frequency, and $\omega$ is the frequency of the incoming radiation. As it was shown in Ref. 34, the resonant response in the presence of a drain current can be written as in Eq. 11, but with a replacement $\tau \rightarrow \tau_{eff}$. Here, $1/\tau_{eff}$ is the effective linewidth given by:

$$1/\tau_{eff} = 1/\tau - 2v/L \quad (12)$$

where $v$ is the electron drift velocity. With increasing current, the electron drift velocity increases, leading to the increase of $\tau_{eff}$ and of the quality factor. When $\omega_0\tau_{eff}$ approaches unity, the detection becomes resonant.

First experiments that demonstrated the influence of the current on the resonant detection were performed with a 250 nm gate length commercial GaAs HEMT and 200 GHz, 600 GHz sources as well as femtosecond THz sources [35-37]. A resonance peak in the detection signal appeared when the transistor was driven into the saturation region.

The effect of the current-driven line narrowing is still more visible in the case of the narrow multichannel transistors. This is illustrated in Fig. 7a, curve 3 (triangles) which was measured in the current-driven detection regime with the applied drain-to-source voltage $V_d = 100$ mV. Comparison of the curves 2 and 3 shows that the linewidth of the resonance clearly decreases when current is applied [33]. Systematic studies have shown that the quality factor could be increased by more than an order of magnitude. This effect can be illustrated by plotting the quality factor versus the drain voltage as in Fig. 7b. The continuous line is calculated by using Eq. 13 [34].

$$Q = \omega\tau_{eff} = 2\pi f\left(\frac{\tau L}{L - 2v\tau}\right) \quad (13)$$

The agreement between the experimental data and calculations provides a clear indication that the dc current may significantly increase the quality factor and drive the two-dimensional plasma towards the Dyakonov-Shur plasma wave instability.

The role of the gate in a FET used as THz detector is not limited to control the carrier density in the channel. In single-gate FETs, the plasma waves under the gate, are only weakly directly coupled to the radiation [38, 39] due to much bigger wavelength of the radiation then dimensions of the transistor (only non direct coupling is provided by the antennas). A conventional way to directly couple efficiently the plasma waves to the incident radiation is to use grating gate coupler structures [40, 41]. In Ref. 41, Shaner *et al.* used an innovative geometry with a split grating-gate and found that only a small portion centered on the pinched-off single finger is active and contributes to the photoresponse. In order to make the active area more effective, doubly interdigitated grating gate detector structures were proposed and their performance as efficient THz sources was verified [42-45]. The first investigation showed that a doubly interdigitated grating gate can enhance the nonresonant photoresponse at room temperature and sub-THz radiation frequencies [45].

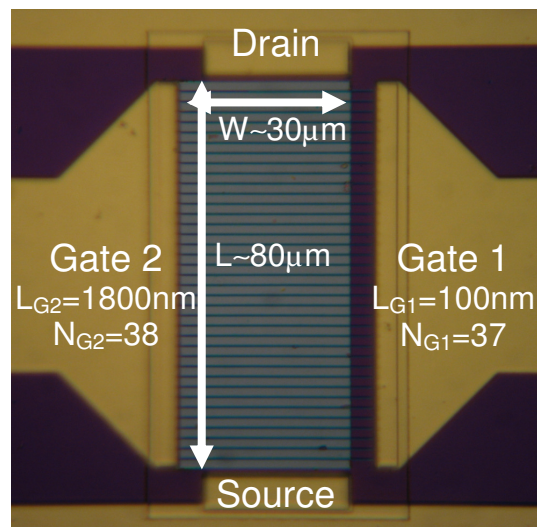

Fig. 8. A scheme of doubly interdigitated grating gate detectors. *L* and *W* are the active area length and width, respectively. One period of the structure consists of one finger of Gate 1, one finger of Gate 2 and two ungated regions (100 nm).

Figure 8 illustrates the top section of one of the grating gate transistors. The device structure is based on an InGaAs HEMT and incorporates doubly interdigitated grating gates G1 and G2 with number of fingers $N_{G1}$ and $N_{G2}$, respectively. The gates periodically modulate the carrier density in the regions underneath.

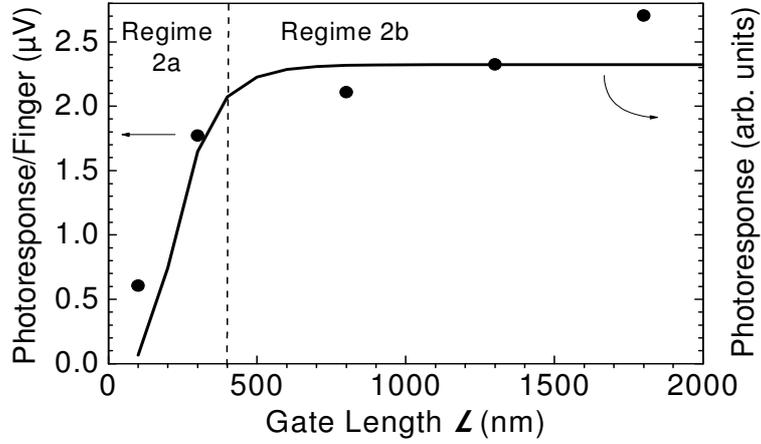

Fig. 9. Measured intensity of the photoresponse per finger (dots) as a function of the active gate finger length (gate at the threshold voltage). The solid line shows the result of calculations of the maximum photoresponse per finger after Ref. 38 versus gate length with $\tau = 0.17$ ps ($\mu = 7000$ cm$^2$/(Vs)). The dotted line schematically separates the regimes 2*a* and 2*b*.

To investigate a non-resonant photoresponse to 238 GHz radiation as a function of the fingers length, the signal was recorded when pinching-off the gates $G_1$ and $G_2$ independently. A gate biased close to the threshold voltage is then active, while a gate at the zero bias voltage is passive. Under the assumption that the photoresponse is the sum of the response of each finger of the active gate, the influence of the finger length can be determined by normalizing the photoresponse near the threshold voltage to the number of fingers of the active gate (dots in Fig. 9). For $f = 238$ GHz ($\omega\tau \sim 0.37$), the normalized photoresponse is almost constant for large gate finger lengths, whereas it decreases with the gate finger length below 400 nm. As shown in Ref. 9, the maximum of the photoresponse near the threshold voltage can be described by a theoretical model that combined the effect of exponential decrease of the electron density and the gate leakage current. The solid line in Fig. 9 shows the calculations of the maximum photoresponse as a function the gate finger length for the relaxation times $\tau = 0.17$ ps, which correspond to $\mu = 7000$ cm$^2$/(Vs). This theory predicts that the maximum of the photoresponse is constant for large gate finger lengths and decreases when this length is below a certain value (dotted line). The dotted line separates the regimes 2*a* and 2*b* discussed in Sect. 2. A reasonable agreement between the experimental results and the theory indicates the validity of the assumption that the nonresonant photoresponse is the sum of the responses of each finger of the active gate.

## 4. FETs for room temperature THz imaging

Since the first demonstrations in 1995 of all-optoelectronic terahertz imaging with a pulsed time-domain setup, various all-optoelectronic systems have been developed, either based on pulsed, continuous-wave (CW), or quasi-CW near-infrared laser sources. None of these approaches achieved real-time imaging capabilities, mostly because of the lack of easily integrable matrices of fast and sensitive detectors.

As already mentioned above, FETs can detect THz radiation at room temperatures in the nonresonant mode. Their nonresonant detection signal ($\omega\tau << 1$) corresponding to the cases 2*a* and 2*b* in Sect. 2, increases monotonically as the gate polarization approaches the threshold. Hence, even if the broadening of plasma resonances (Sect. 3) is still unclear, it is evident that FETs can be used for THz detection. The point is that even in the absence of plasma waves, the gated plasma provides an efficient rectification mechanism useful for fast broadband detection. Obviously, FET transistors are easily integrable in multipixel systems forming focal plane arrays. They can also be efficient when working with various existing sources. Successful tests were performed for THz detection/imaging experiments with optically-pumped molecular THz lasers, time domains spectroscopy (TDS) systems, Gunn diodes, BWO and Quantum Cascade Lasers. Below we give some of the most important results related to a possibility of the room temperature imaging by FETs.

The first imaging experiments with a GaAs/AlGaAs FET acting as a detector were performed using a single pixel/transistor system operating at room temperature. The experiments were performed at 0.6 THz [46]. A focused beam scanned the "Croix du Languedoc", concealed in a paper envelope and attached to a *xy* linear motor stage. The power of the incident beam on the cross was 90 µW, the spot radius was 700 µm, and the response of the transistor was measured using a lock-in amplifier with an integration time of 10 ms. For comparison, a detector array operating at video rates would require pixel elements with 30 ms response time. Thus, this data suggests that a transistor array could be an excellent solution for a real time THz camera.

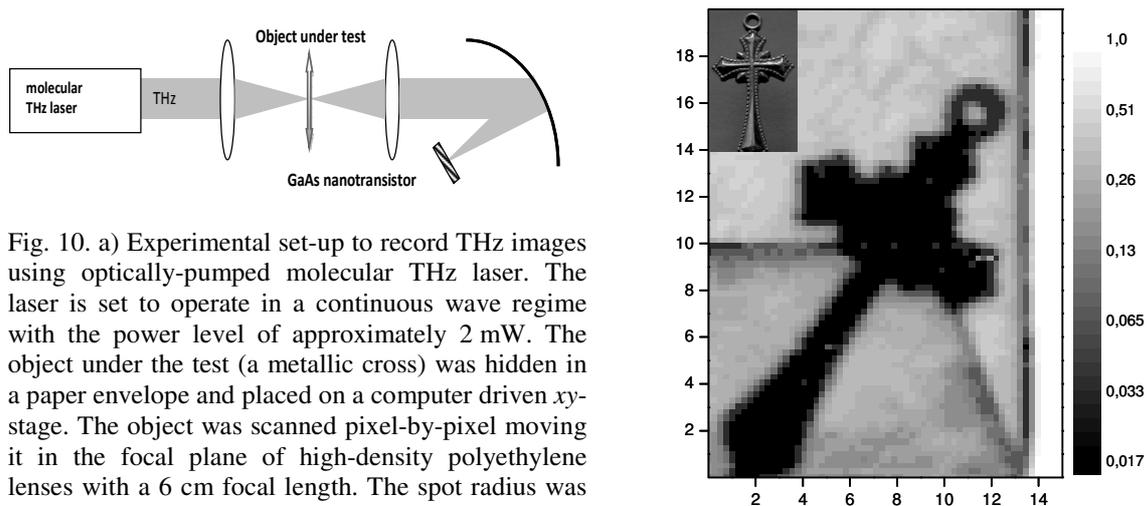

Fig. 10. a) Experimental set-up to record THz images using optically-pumped molecular THz laser. The laser is set to operate in a continuous wave regime with the power level of approximately 2 mW. The object under the test (a metallic cross) was hidden in a paper envelope and placed on a computer driven *xy*-stage. The object was scanned pixel-by-pixel moving it in the focal plane of high-density polyethylene lenses with a 6 cm focal length. The spot radius was determined to be 0.8 mm. The THz light was focused with a spherical mirror (a 11cm focal length) on a GaAs nanotransistor.

Fig. 10. b) The THz image of a metallic cross at 1.63 THz concealed in a paper envelope. Inset shows a photo of the cross for comparison. Scales are in millimeters, linear intensity scale is given in relative units.

The first room temperature imaging experiments using HEMTs as detectors were performed only for sub-THz frequencies. In most experiments, the detected signal decreased dramatically with the increase of the incident radiation frequencies either because of a strong reduction in coupling efficiency and/or because of the water vapor absorption. Only recently, the first images taken in a transmission mode at 1.63 THz using GaAs HEMTs at room temperature were demonstrated [47]. The recording set-up was based on optically-pumped molecular THz laser (see Fig. 10a). The device operated in a photovoltaic mode, without application of the current through the channel, at a gate voltage of -0.45 V. As one can see from the image consisting of 60×80 pixels (Fig. 10b), the nanotransistor resolves a metallic cross (see inset). Note, that the sticker part of the envelope is also clearly distinguished, even the structure of the slight paper bent is visible. Smooth shadowed areas over the entire image occur due to the reflection between the envelope sheets. Signal-to-noise ratio was determined to be about 100:1, with the integration time of 200 ms.

Above we presented results obtained with continuous wave sources. A natural question is whether FET-based detectors can also work with time-domain spectroscopy (TDS) systems. Teppe *et al.* have shown [36] that GaAs FETs can be used for TDS once their sensitivity is enhanced by the passing a drain current. They reported on a room-temperature, resonant detection of femtosecond pulsed THz radiation obtained by optical rectification in a ZnTe crystal. The laser used in this experiment was a Ti/Sapphire amplifier Coherent Hurricane with 800 nm central wavelengt, 130 fs pulse duration, and 1 kHz repetition rate. The maximum amplitude of the emitted THz spectrum was at 0.64 THz and its width was 0.9 THz. The detection was realized using a 250 nm gate length GaAs/AlGaAs FET. The detection amplitude was strongly enhanced (two orders of magnitude) by increasing the drain current and driving the transistor close to the saturation region. These first results have clearly shown that FETs can be efficiently used as fast detectors for THz spectroscopic imaging based on the femtosecond pulsed THz sources.

Very recently, first THz imaging system using pulsed time-domain setup with the resonant detection by plasma waves was reported [48]. Using a GaAs HEMT working with a constant drain bias, the authors obtained high quality images presented in Fig. 11. The results of a raster-scan imaging in the transmission mode of a metallic paper clip, with and without an envelope are presented. The system was based on a femtosecond Ti:Sapphire laser oscillator, it was relatively compact and simple since it did not require a pump-probe time delay. Consequently, the complexity of the optics could be greatly reduced.

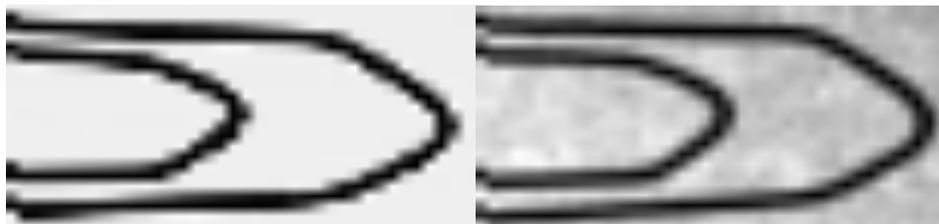

Fig. 11. Raster-scan imaging in transmission mode of a metallic paper clip. Left: without envelope, right: in envelope. 20 mm x 10 mm THz image was made with a numerical aperture of 0.5 and 0.3 mm pixel size (after Ref. 48).

From the point of view of applications, the most promising are the results on THz detection based on FETs made in the silicon technology [13, 14]. This is because they indicate directly the possibility of low cost focal plane array systems for THz imaging [15, 16]. Fig. 12 shows the noise equivalent power (NEP) measured at room temperature at 0.7 THz with a single nanometer-sized Si MOSFET. One can see that the value of NEP is in the $10^{-10}$ W/Hz$^{1/2}$ range which is comparable to the best current commercial room temperature THz detectors.

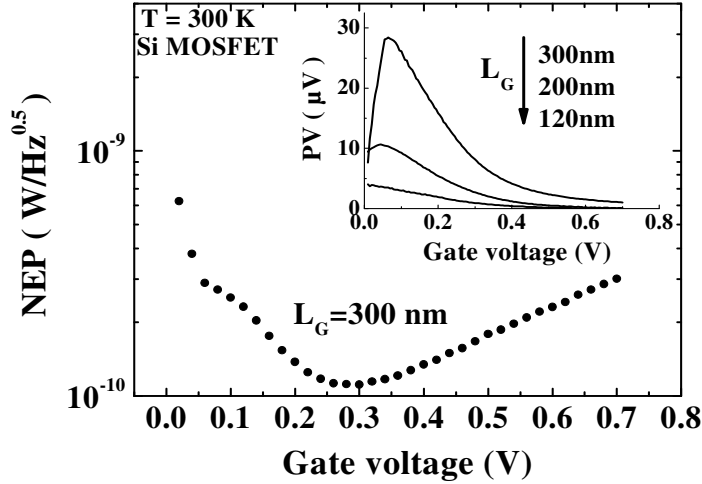

Fig. 12. NEP as a function of the gate voltage for Si MOSFETs with a 300 nm gate length, $T = 300$ K. The inset shows the detection signal as a function of the gate length (after Ref. 14).

The inset of Fig. 12 shows the detection signal as a function of the gate length. One can see that the detected signal decreases when the gate length is reduced from 300 nm to 130 nm. This can be easily explained by using the considerations of Sect. 2 and Fig. 2. In these experiments, the transistor is working in the regime in which by decreasing the transistor length below the damping length $l$ (passing from the case 2*b* to 2*a*) one looses a part of the photoresponse.

A very important progress has been reported recently: an efficient 4-multipixel detection with Si MOSFET devices [15]. The authors confirm the results of Tauk *et al.* [14] and show that the Si-CMOS technology can give NEP in the 100 pW/Hz$^{1/2}$ range. An amplifier was integrated within each pixel to increase the sensitivity up to kV/W. Multipixel reading allowed to increase the speed of collecting images, clearly demonstrating the possibility of imaging at a video rate. Another important work has been reported very recently: a CMOS focal-plane array for heterodyne terahertz imaging [49]. In this work a 3×5 pixel array was fully integrated on a chip and consisted of differential patch antennas, NMOS square-law mixers, and 43 dB low-IF amplifiers. The performance of the receiver is comparable with conventional detectors of THz radiation [16] with the advantage of a low-cost and a multi-pixel integration capability offered by the CMOS technology.

## 5. Discussion and conclusions

It has been shown that THz detection by nanometer size FETs persists from cryogenic up to room temperatures and that detection becomes efficient enough to make nanotransistor-based detectors operational for THz imaging applications.

In spite of a well developed theoretical basis and many experimental observations, various basic physics and engineering aspects of THz detection by FETs are still not well understood. The open questions are: i) what is the main plasma resonance broadening mechanism and is it possible to have a resonant detection at room temperature and ii) what is the best antenna/grating structure to make transistors operating also at higher THz frequencies. A partial solution of the first problem was obtained by application of the drain current, which results in a line narrowing and pushes the system towards the Dyakonov-Shur instability. However, efficient room temperature tunability is still to be demonstrated. The second main problem, concerning the efficiency of the radiation coupling, was already partially solved for sub-THz range by use of integrated antenna [15, 26]. For higher frequencies grating structures could be a good alternative [42].

The physical basis of the THz detection presented above shows that the photovoltaic detection signal is due to a rectification of the incident radiation by plasma nonlinearities. The most important one is caused by the modulation of the electron density by the local gate voltage. The rectification mechanism can be modeled by an RLC line in the resonant case and an RC line in the nonresonant one [31]. The operation regime depends not only on the quality factor but also on the relation between the gate length and the characteristic damping length.

The experimental and theoretical results clearly indicate that nanometer transistors are promising candidates for a new class of efficient THz detectors. The natural next step is the realization of real-time imaging THz cameras. To understand whether FETs are the best candidates for this purpose, let us briefly consider other approaches that have already demonstrated their potential in THz real-time recording systems.

The simplest way is using a commercial infrared 160×120 element microbolometer camera. Although the device is designed for wavelengths of 7.5-14 μm, it retains the sensitivity to the THz radiation delivered by optically-pumped molecular THz laser [50]. It was shown that in a transmission-mode THz images can be obtained at the video rate of 60 frames/s; signal-to-noise ratio is estimated to be 13 dB for a single frame of video at 10 mW power. An essential step in scaling down the dimensions of a real-time imaging system is the replacement of the optically-pumped laser by a quantum cascade laser [51]. For instance, a quantum cascade laser operating at 4.3 THz with the power of 50 mW allowed reaching the signal-to-noise ratio of 340 at 20 frames/s acquisition rate and an optical NEP of 320 pW.

In another promising approach based on a thin-film absorber upon a silicon nitride membrane, with thermopile temperature readout produced with the CMOS technology [52], a 5 ms thermal time constant of the detector, together with the noise equivalent power of 1 nW/Hz$^{1/2}$ enables the real-time imaging at 50 frames/s with a signal-to-noise ratio of 10 for an optical intensity of 30 μW/cm$^2$. Very recently, THz images below 1 THz at room temperature were recorded using InGaAs-based bow-ties diodes with a broken symmetry [53]. The operation principles rely on a non-uniform carrier heating in a specific diode structure merging an antenna concept for coupling of the radiation and a high mobility 2DEG as an active medium [54]. The response time was found to be less than 7 ns, the NEP of about 5.8 nV/√Hz, the sensitivity in the range of 6 V/W, and the dynamic range of about 20 dB at the bandwidth of 100 MHz.

In this context, FETs can be regarded as the most promising option, in particular, Si-based devices displaying a high responsitivity, NEP close to 100pW/√Hz [14] and possibility of integration in the focal plane arrays [15,16,49].

### Acknowledgements

We thank prof. T. Skotnicki (ST Microelectronics) for providing the Silicon FETs, prof. A. Cappy and prof. S. Bollaert (IEMN, Lille) for providing InGaAs HEMTs. We thank also dr. P. Mounaix and dr. E. Abraham (LPMOH CNRS and Bordeaux I University) for their experimental support in the time domain spectroscopy. This work was financially supported in part by JSPS International Fellowship Program for Research in Japan, by the joint French-Lithuanian research program "Gilibert/EGIDE.", and by the joint French-Japanese research program "Sakura/EGIDE.". JŁ, WK and KK acknowledge the support of 162/THz/2006/02 and MTKD-CT-2005-029671 grants. The authors from the Montpellier University acknowledge the CNRS guiding GDR and GDR-E projects "Semiconductor sources and detectors of THz frequencies" and the Region of Languedoc-Roussillon through the "Terahertz Platform" project, as well as ANR TeraGaN project. Experiments at Vilnius were conducted under the project "Terahertz optoelectronics: devices and applications" (No. 179J).